\documentclass[submission, Proceedings]{SciPost}

\begin{document}

\begin{center}{\Large \textbf{
$\tau$ lepton mass measurement at BESIII
}}\end{center}

\begin{center}
J. Y. Zhang\textsuperscript{1*} on behalf of BESIII collaboration
\end{center}

\begin{center}
{\bf 1} Institute of High Energy Physics, Chinese Academy of Sciences, Beijing 100049, China
\\
* jyzhang@ihep.ac.cn
\end{center}

\begin{center}
\today
\end{center}

\definecolor{palegray}{gray}{0.95}
\begin{center}
\colorbox{palegray}{
  \begin{tabular}{rr}
  \begin{minipage}{0.05\textwidth}
    \includegraphics[width=8mm]{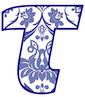}
  \end{minipage}
  &
  \begin{minipage}{0.82\textwidth}
    \begin{center}
    {\it Proceedings for the 15th International Workshop on Tau Lepton Physics,}\\
    {\it Amsterdam, The Netherlands, 24-28 September 2018} \\
    \href{https://scipost.org/SciPostPhysProc.1}{\small \sf scipost.org/SciPostPhysProc.Tau2018}\\
    \end{center}
  \end{minipage}
\end{tabular}
}
\end{center}


\section*{Abstract}
{\bf
In order to obtain the mass of $\tau$ lepton at BESIII precisely, a beam energy measurement system was built at BEPCII. A scenario for high precision $\tau$ mass measurement was put forth before data taken. More than 130 $pb^{-1}$ $\tau$ mass scan data were collected in April 2018, and the uncertainty of $m_{\tau}$ is expected to be less than 100 keV.
}

\vspace{10pt}
\noindent\rule{\textwidth}{1pt}
\tableofcontents\thispagestyle{fancy}
\noindent\rule{\textwidth}{1pt}
\vspace{10pt}

\section{Introduction}
\label{sec:intro}
The $\tau$ lepton is a fundamental particle in the standard model, and its mass is a fundamental parameter and should be given by experiment accurately. According to PDG~\cite{pdg2018}, the current world average value of the $\tau$ lepton mass is $m_{\tau}^{PDG} = 1776.86 \pm 0.12$ MeV. It is based mostly on BESIII~\cite{bes32014}, KEDR~\cite{kedr}, BES~\cite{bes}, BABAR~\cite{babar} and BELL~\cite{bell}. The latter two experiments are performed at B factory using the pseudomass method, analysing the huge amount of data, they obtained good statistical accuracy, however the systematic uncertainty is large for the absolute calibration of particle momentum measurements. The former three experiments obtained their results by scanning the $\tau$  threshold region. The $\tau$ mass value was extracted from the dependence of the production cross section on the beam energy. Near the threshold, the luminosity of data is limit, and the statistical uncertainty is large. The systematical uncertainty related on the energy scale are dominant. In order to decrease the systematical uncertainty related on the energy scale, beam energy measurement system was built at BEPCII.

The paper is organized as follows: the beam energy measurement system is introduced in section~\ref{sec:bems}. Monte carlo simulation for high precision $\tau$ mass measurement was described in section~\ref{sec:opt}.  In the spring of 2018, we took some scan data near $\tau$ threshold, some analysis work and statistic uncertainty estimation are introduced in section~\ref{sec:esti}, then we will give a short summary.


\section{Beam energy measurement system at BEPCII}
\label{sec:bems}
The beam energy measurement system (BEMS) at BEPCII is based on the Compton backscattering (CBS) principle. The working scheme of this system is as follows~\cite{nickolai, principle}: a laser source provides a laser beam, and an optics system focuses the laser beam and guides it to collide with the electron (or positron) beam in the vacuum pipe, where the CBS process happens; after that the backscattering high energy photons are detected by a HPGe detector.

\begin{figure}
\begin{center}
\begin{minipage}{12cm}
\includegraphics[width=12cm]{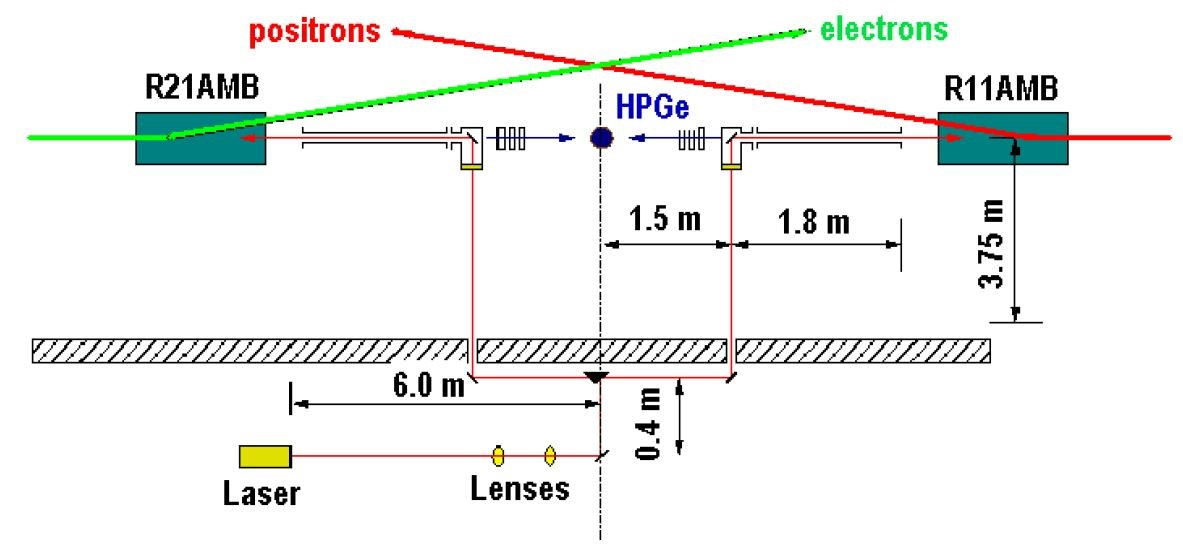}
\end{minipage}
\caption{
\label{bems}
Simplified schematic of BEMS. The positron and electron beams are indicated. R1IAMB and R2IAMB are accelerator magnets, and the HPGe detector is represented by the dot at the center. The shielding wall of the beam tunnel is shown cross-hatched, and the laser is located out-side the tunnel.
}
\end{center}
\end{figure}
The layout schematic of the system is shown in Fig.~\ref{bems}. According to CBS theory, the backscattering photon, $\omega_{max}$, has relation with the beam energy $\varepsilon$~\cite{rus, moxh}:
\begin{equation}
\omega_{max} = \frac{\varepsilon^{2}}{\varepsilon + m^{2}_{e}/4\omega_{0}}~,
\label{omg}
\end{equation}
where $\omega_{0}$ = 0.117065228 eV,  is the energy of initial photon emitted by the laser, the photon is produced by a GEM selected 50$^{TM}$ continuous wave CO$_{2}$ laser. $\omega_{max}$ can be determined through the detection of scattered photons by HPGe detector. Then the beam energy can be deduced from the following formula,
\begin{equation}
\varepsilon = \frac{\omega_{max}}{2} \left[1+\sqrt{1+\frac{m^{2}_{e}}{\omega_{0}\omega_{max}}}\right]~.
\label{nepec}
\end{equation}

The HPGe detector we used in our experiment is a coaxial n type germanium detector manufactured by ORTEC, whose model is GMX25P4-76. It has diameter of 58.9 mm and height of 55.0 mm with relative efficiency of 35.2\%. The energy resolution for the 1.33 MeV line of $^{60}$Co is 1.90 keV (FWHM). The detector is connected to the digital signal processing unit Dspec Pro, the integral and differential nonlinearities are $\pm$ 250 ppm and $\pm$ 1\% respectively.

In order to describe the asymmetry of detection resolution, an asymmetry response function for HPGe detector was used as follows:
\begin{equation}
f(x) = \frac{\sqrt{2/\pi}}{\sigma_{R}+\sigma_{L}} \left\{
\begin{array}{cc}
exp(-x^{2}/2\sigma_{R}^{2})  &  {\rm if} ~~~~ x >= 0; \\
exp(-x^{2}/2\sigma_{L}^{2})  &  {\rm if} ~~~~ x < 0. \\
\end{array}
\right.
\label{resps}
\end{equation}
Where x = E - $E_{max}$ is the difference between the energy detected by HPGe detector and the most probable value. $\sigma_{R}$ and $\sigma_{L}$ represent the energy resolution of detector on the right side and the left side. The energy is derived from multichannel analyzer by means of the linear transformation:
\begin{equation}
E_{\gamma} [keV] = zero [keV] + gain [keV] \times channel,
\label{linear}
\end{equation}
where, zero and gain are the calibration coefficients, their unit is keV.

The detector is calibrated in real time during data taking. The calibration sources generally used in experiments are listed in Table~\ref{lst} below:
\begin{table}[htb]
\caption{\label{lst}The list of calibration lines. }
\begin{center}
\small
\begin{tabular}{|c|c|} \hline
source&$\gamma$-rays energy, keV  \\ \hline
$^{137}$Cs & 661.657  $\pm$ 0.003 \\ \hline
$^{60}$Co & 1173.228  $\pm$ 0.003 \\ \hline
$^{60}$Co & 1332.492  $\pm$ 0.004 \\ \hline
$^{208}$Tl & 583.187  $\pm$ 0.002 \\ \hline
$^{208}$Tl & 2614.511  $\pm$ 0.010 \\ \hline
\end{tabular}
\end{center}
\label{wellknown}
\end{table}

In our experiment, the energy region is from keV to several MeV. In order to reduce the influence of electronics nonlinearity to our measurement, a precise pulse generator with model PB-5 is used, whose integral nonlinearity is $\pm$ 15 ppm~\cite{bnc}. The output of the generator is connect to the preamplifier of HPGe detector, a voltage signal is input to the detector, a energy signal is obtained by multichannel analysis.


The absolute energy scale is determined as follows: first, linear scale calibration with radiation sources. We adjust the gain in Eq.~\ref{linear} and redefine the horizontal axis of the histogram accordingly. When the linear fit of $E_{FIT}$ - $E_{REF}$ vs $E_{\gamma}$ equals zero, the zero and gain is defined, where the $E_{FIT}$ - $E_{REF}$ is the discrepancies between the peak energies obtained from fits and the corresponding reference energies from table~\ref{wellknown}, $E_{\gamma}$ is the $E_{FIT}$ as the red dots shown in Fig..  Second, correction of the electronics nonlinearity by precision pulser signals. A set of pulser lines are generated in the spectrum as the green triangles shown in Fig.. A linear conversion is used to describe the pulser amplitudes $A_{i}$ to corresponding energies: 
\begin{equation}
E^{i}_{REF} [keV] = P_{0} [keV] + P_{1} [keV/V]  \times A_{i} [V]. 
\end{equation}
All of the pulser peaks in the spectrum are fitted, their energy are assigned to be $E^{i}_{FIT}$.
The dependence ($E^{i}_{FIT}$ - $E^(i)_{REF}$) vs $E_{\gamma}$ = $E^{i}_{FIT}$ is fitted by the univariate spline. Then we adjust the parameters $P_{0}$ and $P_{1}$ to minimize the difference between this line and the points of absolute calibration, The energy scale is calibrated in the whole energy region.

\begin{figure}
\begin{center}
\begin{minipage}{12cm}
\includegraphics[height=6cm,width=12.cm]{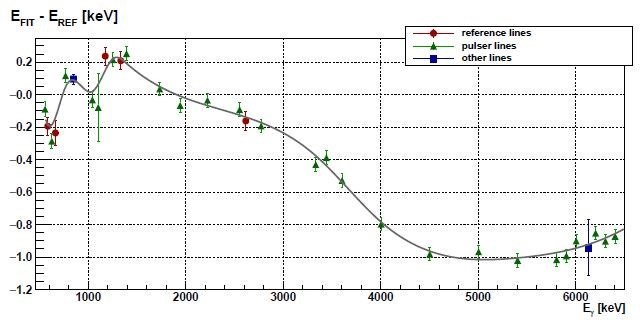}
\end{minipage}
\caption{
\label{calib}
The energy scale calibration for HPGe detector. The red circles are obtained from the radiation sources, the
green triangles are from pulser peaks in the spectrum, the blue squares are used for calibration check.}
\end{center}
\end{figure}

The above procedure is based on the assumption that all the integral nonlinearity in energy determination is caused by electronics. Two known energy lines a applied to verify the accuracy of this assumption, they are $^{56}Mn$ and $^{16}O$ as the blue square shown in Fig.~\ref{calib}.  Both of them are agree well with the calibration line.

After good calibration, the BEMS will provide energy value for beam precisely.

\section{Optimization of data taken scenario}
\label{sec:opt}
As mentioned above, the threshold scan is adopted by BESIII experiment to determine the mass of $\tau$ lepton. This method is dependent on the behavior near threshold. The expected observed cross section can be written as:

\begin{equation}
\sigma^{obs}(W,m_{\tau},\epsilon,\sigma_{B}) = \epsilon \times \sigma(W) + \sigma_{B}
\end{equation}
where $\epsilon$, $\sigma_{B}$, and W are the overall detection efficiency, background cross section, and C.M. energy respectively. In order to achieve the highest possible accuracy of the mass measurement, the optimization of luminosity and location of the energy points is necessary.

In our experiment, three parameters of $m_{\tau}$, $\epsilon$, and $\sigma_{B}$ need to be fit.
It is necessary to have at least three energy points to obtain the three parameters of the fit. For the three points scenario, the only constraints required for optimization are the total integrated luminosity and the full energy range, which must be narrow enough to ensure a sufficient uniformity of the background and to minimize the efficiency variation.

\begin{figure}
\begin{center}
\begin{minipage}{12cm}
\includegraphics[height=6cm,width=12.cm]{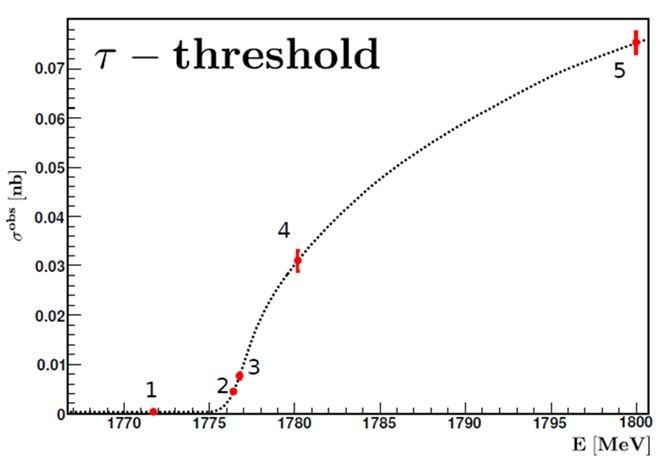}
\end{minipage}
\caption{\label{opt}
The dependence of observed cross section on the beam. The dots with error bar shows the expected measurements using the tau threshold scan scenario.}
\end{center}
\end{figure}

The three point scenario~\cite{wangyk} does not give information about possible instabilities during the scan (detector efficiency variation, beam instabilities, energy measurement instabilities) and is sensitive to the uncertainty of the PDG tau mass value. Additional points must be added to check the threshold shape. Therefore we need five points scenario~\cite{ivan} as shown in Fig.~\ref{opt}.

The first point is to determine the background. Two points (P2 and p3) are located near the tau threshold, the difference of their energy $E_{3}$ - $E_{2}$ = 2.5 $\sigma_{m_{\tau}}^{PDG}$,  where $\sigma_{m_{\tau}}^{PDG}$ is the PDG error of the $\tau$ mass. This constraint reduces the sensitivity to
the assumed mass value. The luminosity fractions of p2 and p3 should be similar since they are close to the
uncertain  tau mass value. We choose  $L_{2}$:$L_{3}$ = 3:2. The fourth point (p4) should be higher than the tau threshold, its purpose is to check the cross section shape, less luminosity is needed. The last point (p5) is at the high energy region, it can determine the detection efficiency. The luminosity ratio $L_{4}$:$L_{5}$ = 1:2.

\section{Estimation of statistical uncertainty for $\tau$ mass}\label{estimation}
\label{sec:esti}
In the spring 2018, BESIII detector took tau threshold scan data, BEMS played an important role to determine the beam energy precisely. At first, we scan $J/\psi$ resonance with 7 points; then, performed the $\tau$ threshold scan using five points; at last, performed the $\psi(2S)$ scan with 9 points. In order to decrease the uncertainty, the statistic uncertainty of each point is about 0.1 MeV. Total about 130 pb$^{-1}$ $\tau$ data were collected.

When the raw data are reconstructed, the data analysis work is performed. We focus on the e $\mu$ and e $\pi$ final state modes. According to our preliminary analysis, the statistic uncertainty of the $m_{\tau}$ is about 70 keV. If we compare this result with the analysis of $\tau$ scan performed in 2011, extend the final state decay mode to 13,
the statistical uncertainty will be less than 45 keV. If the total uncertainty is required to be less than 0.1 MeV,
the systematical uncertainty will be 90 keV.

\section{Conclusion}
\label{sec:con}
In order to determine the mass of $\tau$ lepton precisely, beam energy measurement system was built at BEPCII, and commissioning is well. Monte Carlo simulation was performed to optimization the position of scan points and the luminosity allocation. $\tau$ threshold scan was performed at BESIII this spring, more than 130 $pb^{-1}$ data were collected. Data analysis on statistic and systematic uncertainty are in progress, the total uncertainty of mass of $\tau$ lepton is expected to be less than 0.1 MeV.







\section*{Acknowledgements}


\paragraph{Funding information}
This work is supported in part by National Natural Science Foundation of China (NSFC)
under contracts Nos.: 11375206, 10775142, 10825524, 11125525, 11235011, Y81161005Z; the
Ministry of Science and Technology of China under Contract Nos.: 2015CB856700,
2015CB856705; State key laboratory of particle and detection and electronics;
and the CAS Center for Excellence in Particle Physics (CCEPP);
the RFBR grant No 14-02-00129-a; U.S. Department of Energy  under  Contracts  No.  DE-FG02-04ER41291, No. DE-FG02-05ER41374, No. DE-FG02-94ER40823, No. DESC0010118; U.S. National Science Foundation; RFBR 18-52-53014-NSFC-a; part of this work related to the data analysis is supported by the Russian Science Foundation (project No 14-50-00080)







\bibliography{SciPost_Example_BiBTeX_File.bib}

\nolinenumbers

\end{document}